\documentclass[twocolumn,showpacs,preprintnumbers,amsmath,amssymb]{revtex4}

\usepackage{bm}
\usepackage{amsthm}
\usepackage{amssymb}
\usepackage{amsfonts}
\usepackage[dvips]{graphicx}
\usepackage{subfigure}
\usepackage[english]{babel}
\usepackage{amssymb}
\usepackage{indentfirst}
\usepackage{graphics}
\usepackage{amsmath}
\usepackage{multirow}

\newcommand{\beq}{\begin{equation}}
\newcommand{\eeq}{\end{equation}}
\newcommand{\bcen}{\begin{center}}
\newcommand{\ecen}{\end{center}}
\newcommand{\bdm}{\begin{displaymath}}
\newcommand{\edm}{\end{displaymath}}
\newcommand{\best}{\begin{equation*}}
\newcommand{\eest}{\end{equation*}}
\newcommand{\bea}{\begin{eqnarray}}
\newcommand{\eea}{\end{eqnarray}}

\newcommand{\wid}{\textrm{wd}}

\newcommand{\out}{\textrm{out}}
\newcommand{\locc}{\textrm{LOCC}}
\newcommand{\poly}{\textrm{poly}}

\newcommand{\Ee}{E_\epsilon}

\newcommand{\bra}[1]{\langle #1 |}
\newcommand{\ket}[1]{| #1 \rangle}

\newcommand{\pro}[1]{\ket{#1}\bra{#1}}
\newcommand{\proj}{\pro}

\newcommand{\sphi}{|\varphi\rangle}
\newcommand{\sphii}{|\varphi_i\rangle}

\newcommand{\spsi}{|\psi\rangle}

\newcommand{\spsii}{|\psi_i\rangle}

\renewcommand{\epsilon}{\varepsilon}
\renewcommand{\phi}{\varphi}

\newcommand{\tr}{{\textrm{Tr}}}

\newcommand{\mco}{{\mathcal{O}}}

\newcommand{\mcs}{{\mathcal{S}}}

\newcommand{\naturali}{{\mathbb{N}}}

\theoremstyle{plain}
\newtheorem{prop}{Proposition}
\newtheorem{obs}{Observation}
\newtheorem{cor}{Corollary}
\newtheorem{???}{Question}
\newtheorem{defi}{Definition}
\newtheorem{teo}{Theorem}
\newtheorem*{teo*}{Theorem}
\newtheorem{lemma}{Lemma}
\newtheorem{dim*}{Proof}

\newtheorem{reason*}{Motivation for the question}

\theoremstyle{definition}
\newtheorem{es}{Example}

\theoremstyle{remark}

\begin{document}

\title{Universal resources for approximate and stochastic measurement-based quantum computation}
\author{Caterina E. Mora$^{1,3}$, Marco Piani$^{2,3}$,
Akimasa Miyake$^{1,2,4}$, \\
Maarten Van den Nest$^{1,5}$, Wolfgang D{\"u}r$^{1,2}$ and Hans J. Briegel$^{1,2}$}

\affiliation{
$^1$Institut f{\"u}r Quantenoptik und Quanteninformation der \"Osterreichischen Akademie der Wissenschaften, Innsbruck, Austria \\
$^2$Institut f{\"u}r Theoretische Physik, Universit{\"a}t Innsbruck, Technikerstra{\ss}e 25, A-6020 Innsbruck, Austria \\
$^3$Institute for Quantum Computing \& Department of Physics and Astronomy,
University of Waterloo, University Ave. W., N2L 3G1, Canada \\
$^4$Perimeter Institute for Theoretical Physics, 31 Caroline St. N., Waterloo ON,
N2L 2Y5, Canada \\
$^5$Max-Planck-Institut f{\"u}r Quantenoptik, Hans-Kopfermann-Str. 1, D-85748 Garching, Germany }

\date{\today}

\begin{abstract}
We investigate which quantum states can serve as {\em universal resources} for approximate and stochastic measurement-based quantum computation, in the sense that any quantum state can be generated from a given resource by means of single-qubit (local) operations assisted by classical communication. More precisely, we consider the approximate and stochastic generation of states, resulting e.g. from a restriction to finite measurement settings or from possible imperfections in the resources or local operations. We show that entanglement-based criteria for  universality obtained in [Van den Nest et al, New J. Phys. 9, 204 (2007)] for the exact, deterministic case can be lifted to the much more general  approximate, stochastic case. This allows us to move from the idealized situation (exact, deterministic universality) considered in previous works, to the practically relevant context of non-perfect state preparation.

We find that any entanglement measure fulfilling some basic requirements needs to reach its maximum value on  some element  of an approximate, stochastic universal family of resource states, as the resource size grows. This allows us to rule out various families of  states as being approximate, stochastic universal. We prove that approximate, stochastic universality is in general a weaker requirement than deterministic, exact universality and provide resources that are efficient approximate universal, but not exact deterministic universal.

We also study the robustness of universal resources for measurement-based quantum computation under realistic assumptions about the (imperfect) generation and manipulation of entangled states, giving an explicit expression for the impact that errors made in the preparation of the resource have on the possibility to use it for universal approximate and stochastic state preparation.

Finally, we discuss the relation between our entanglement-based criteria and recent results regarding the uselessness of states with a high degree of geometric entanglement as universal resources [D. Gross et al., Phys. Rev. Lett. 102, 190501 (2009); M. J. Bremner et al., Phys. Rev. Lett 102, 190502 (2009)].

\end{abstract}


\maketitle

\tableofcontents


\section{Introduction}

Quantum computation offers a new and exciting perspective to  information processing, as it has been found that certain problems can be solved more efficiently on a quantum computer than on a classical device. Despite considerable effort it is however not fully understood which features of quantum mechanics are responsible for the apparent speedup. Basic questions regarding the nature and power of quantum computation remain largely unanswered to date. The existence of various models for quantum computation, among them the quantum Turing machine \cite{Deu85,quantumTM}, the circuit model \cite{BBC+95,CircuitDeutsch,Yao93}, adiabatic quantum computation \cite{FGG+01,latorre} and measurement-based quantum computation \cite{1-way1,1way1long,RBB03,GC99,Leu04,Ni03,PJ04,gross,BM08}, seems to indicate that a straightforward answer to these fundamental issues might be difficult to obtain.

On the other hand, the different nature of the models allows one to study these fundamental issues from different perspectives, and it turns out that some models are better suited than others to study a certain aspect. For instance, the model of {\em measurement-based quantum computation}, with the one-way model \cite{1-way1} as most prominent representative, seems to be particularly well suited to investigate the role of entanglement in quantum computation. Such an investigation has been initiated in \cite{maarten} and further developed in \cite{universalityI}.
 In one-way or measurement-based quantum computation (MQC) --which we use synonymously throughout this article-- a highly entangled resource state, e.g. the 2D cluster state \cite{cluster1}, is processed by sequences of single-qubit local measurements. As has been shown in \cite{1way1long}, a proper choice of measurement directions allows one to generate --up to irrelevant local unitary correction operation-- {\em any} quantum state deterministically and exactly on the unmeasured qubits.  In this paper we aim at investigating the generalization of these previous results to the case in which stochastic and/or approximate quantum computation is allowed.

 The 2D cluster state is called a {\em universal resource} for MQC. In MQC, the role of entanglement is particularly highlighted, as all entanglement required in the computation already needs to be present in the initial resource state. This derives from the fact that no entanglement measure increases under local operations and classical communication (LOCC). This insight was recently used in \cite{maarten,universalityI} to investigate which other quantum states are universal resources for MQC. Entanglement-based criteria for universality have been established and many --otherwise highly entangled-- resource states, including GHZ states \cite{GHZstates}, W states \cite{DurVC00-wstate} and 1D cluster states \cite{cluster1},  have been shown to be not universal for MQC. One should, however, emphasize that this does not mean that such non-universal resource states are useless for quantum information processing, as they might still serve to perform some specific quantum computation or as a resource for some other task. On the positive side, several other states have been identified to be universal resources for MQC \cite{gross,universalityI}. Notice that we use the term ``universality'' in its strongest form, i.e. we consider the generation of quantum states (universal state preparator). This has been termed CQ-universality (where CQ stands for classical input, quantum output) in Ref. \cite{universalityI} and we refer the interested reader to said work for an extended discussion on the different notions of universality.

\subsection{Approximate and stochastic universality}
In this article we will extend the results on universality obtained in \cite{maarten,universalityI} to a more general and realistic setting, which is motivated by experimental reality. More precisely, we will consider the {\em approximate} and {\em probabilistic} generation of quantum states from a given resource state, in contrast to the exact and deterministic generation discussed in \cite{maarten,universalityI}. In this work we therefore focus on the case in which the desired output states are required to be generated only with finite accuracy (that is the output of the computation is required to be within some distance $\epsilon$ of the desired state), and with probability $1-\delta$. Such an extension needs to be considered naturally whenever the resource states are noisy, e.g. due to an imperfect generation process or due to decoherence, but also if the local operations used to process the state are imperfect. The latter may again be reflected in noisy single qubit operations, but may also result from a restriction to a finite number of measurement settings or local unitary operations. In all these cases, the resulting states can only be an approximation of the desired state.

In addition, one might be interested in the generation of states with a probability of success (arbitrary) close to one --which we will call quasi-deterministic--, or even only with some (arbitrary) small success probability. In fact, similar issues are implicitly considered when one refers to universal gate sets in the circuit model for quantum computation: any finite universal gate set  allows one to approximate any state with arbitrary accuracy. Notice that the issue of probabilistic computation has been deeply studied both in classical computation theory \cite{papadim} and in the quantum setting \cite{nielsenchuang}. On the one hand, if it is known when the computation succeeded, which happens, say, with probability $p$, then $O(1/p)$ repetitions allow one to obtain a valid, confirmed outcome. On the other hand, even if it is not known whether the computation succeeded or not, but only that the correct outcome is obtained with some probability $p>1/2$, this is still sufficient to extract the correct (classical) result of the computation with arbitrary high probability by considering many repetitions. The first scenario also applies without changes to the case where quantum states should be generated (CQ universality). The second scenario is restricted to the extraction of classical outputs (CC universality), while the resulting quantum states are in fact mixed.

\subsection{Summary of results}
We find that --analogously to the exact, deterministic case-- entanglement based criteria for approximate and stochastic  universality can be obtained. To formulate these criteria, we need to consider $\epsilon$-measures of entanglement \cite{epsilon} and compute their extremal values over all states. Given any distance $D$ on the set of states, and any entanglement measure $E$, the $\epsilon$-measure of a state $E_{\epsilon}(\rho)$ is defined as the minimal amount of entanglement of all states which are $\epsilon$-close (with respect to $D$) to $\rho$, i.e. have a distance smaller than $\epsilon$ to $\rho$.
 We find the following necessary criteria for efficient, approximate stochastic universality:
\begin{itemize}
\item For any entanglement measure $E$ which is a strong extendable entanglement monotone (see below for exact definition),
we have that an approximate, stochastic universal resource $\Sigma$ which allows one to obtain an $\epsilon$-approximation of any state with probability larger than $1-\delta$, must have an amount of entanglement that is larger or equal than $(1-\delta)$ times the maximum of the corresponding $\epsilon$-measure $E_{\epsilon}$ over all 
 states of arbitrary size, $E(\Sigma) \geq (1-\delta) {\rm max}_{\rho}E_\epsilon(\rho)$.
Roughly speaking, this means that any approximate, stochastic resource needs to be maximally entangled with respect to all such entanglement measures. \item If one takes the efficiency of computation into account, we find that for any strong extendable entanglement monotone, the entanglement of the resource states does not only need to reach the maximum value of the corresponding $\epsilon$-measure over all states, but needs to grow sufficiently fast with the system size.
\end{itemize}
These two criteria allow one to rule out a large number of states as being not universal in an approximate an stochastic sense, e.g. GHZ states, W states and 1D cluster states.

On the positive side, we present a number of approximate, quasi-deterministic resource. We find:
\begin{itemize}
\item There exist efficient, approximate quasi-determinist universal resources that are not believed to be exact, deterministic universal. For example, a 2D cluster state where particles are missing with a certain probability is an exact, quasi-deterministic universal resource, while an approximate 2D cluster state is an approximate deterministic universal resource.

\item Any state that is sufficiently close to an approximate stochastic universal resource is still an approximate stochastic universal resource, and the parameters quantifying approximation and stochasticity are quadratically related to the original ones.
\end{itemize}

In particular, this last observation has implications in realistic (experimental) scenarios, where the preparation of the initial entangled states is imperfect. These errors in the preparation procedure still allow for the state to be used for MQC, in the approximate and stochastic scenario. While this might be considered intuititve and results of this type were already known for the 1-way model (where the initial state is a 2D cluster  state) \cite{raussendorfPhD,nielsendawson2005,aliferisleung2006}, in this paper we extend the observation to all approximate stochastic universal resources, computing an explicit expression for the interplay between the different parameters.


\subsection{Guideline through the paper}
The paper is organized as follows. In Section \ref{sec:entanglement} we review some of the basic concepts, related to distance and entanglement measures respectively, which we use in the remaining of the paper. In Section \ref{sec:universality} we recall the definition of universal resources for measurement-based quantum computation, and see how the definition can be generalized to the approximate and stochastic case. In Section \ref{sec:nogo} we first review some of the results found in \cite{universalityI} and then show how they can be generalized in a very natural way obtaining necessary criteria for universal resources in the approximate and stochastic case. In this Section we also show how the issue of efficiency can be included in the analysis, obtaining in this way stronger versions of the above-mentioned criteria. Finally, in Section \ref{sec:examples}, we give some experimentally relevant examples of resources that are approximate deterministic, exact stochastic and approximate stochastic universal, but not exact deterministic universal. In particular, we show that any family of states that is close to a universal family is still approximate stochastic universal. Section \ref{sec:conclusions} summarizes and concludes our results.

\section{Entanglement monotones}
\label{sec:entanglement}

In this section we review some essential features of entanglement monotones which are relevant in the study of universality in MQC.

In Section \ref{subsec:axioms} we review the basic conditions which a function must satisfy in order to be considered an ``entanglement monotone''. Furthermore, we show how these conditions lead to the definitions of different ``types'' of entanglement measures. The distinction between different types of entanglement measures will be necessary to allow for a proper formulation of entanglement-based criteria for approximate and stochastic universality, as we will do in section \ref{sec:nogo}.

In section \ref{subsec:epsilonmeasures} we consider a general class of monotones called ``epsilon-measures''. This class of measures was introduced in \cite{epsilon} in order to study the entanglement in states which are only known up to some approximation. For this reason they are suitable quantities to consider in the study of \emph{approximate} universality.

In Section \ref{subsec:exmeasures}, we focus on two examples of existing entanglement measures, namely the geometric measure and the Schmidt-rank width. We discuss in which sense these quantities are monotones, and we discuss their associated $\epsilon$-measures.

\subsection{Properties of entanglement monotones}
\label{subsec:axioms}

The first examples of entanglement measures were built by first considering a particular application of entanglement (such as, e.g., distillation) and then deriving a quantifier based on such an operation. This approach led to measures that, while naturally having a clear physical interpretation, were often very hard to compute. To evaluate, for example, the entanglement of distillation \cite{BDSW1996} it is necessary to optimize over all purification protocols. A different approach to the problem, that one might define ``axiomatic'', has been proposed in \cite{VPRK1997}. The starting point of this work was the idea that an entanglement measure is some mathematical quantity that should somehow capture the essential features that we associate with entanglement. With this idea in mind, it is possible to identify a set of conditions that must be satisfied by any such measure $E$. The most fundamental of these conditions are:

\begin{itemize}
\item[P1.]{{\it Vanishing on separable states}}: separable states do not contain entanglement, therefore we require that  $E(\sigma_{sep})=0$. \item[P2.]{{\it Monotonicity under LOCC}}: entanglement cannot increase under LOCC, $E(\Lambda_{\textrm{LOCC}}[\rho])\leq E(\rho)$.
\end{itemize}
 Here $\Lambda_{\textrm{LOCC}}$ denotes an LOCC transformation.
Note that property P2 also implies that $E$ is invariant under local unitaries.

Aside from these two postulates, other additional requirements for entanglement measures have been formulated. In particular, the following are among the most commonly found in literature.
\begin{itemize}
\item[P3.] \label{item1} {{\it Convexity}}: $E(p\rho_1+(1-p)\rho_2)\leq pE(\rho_1)+(1-p)E(\rho_2)$. \item[P4.]  \label{item2}{\it Monotonicity on average under LOCC}: this condition is stronger than the monotonicity condition seen above, and is sometimes referred to as \emph{strong monotonicity}. It requires that the following holds true \beq \label{eq:strongmonotonicity} E(\rho)\geq\sum_i p_i E(\rho_i), \eeq where $\rho_i$ are the possible outputs of some LOCC protocol acting on $\rho$, and occur with probability $p_i$. \item[P5.] {\it Trivial extendability:} in this case, one aims at comparing entanglement in states of different system size. The condition of trivial extendability states the following: let $|\psi\rangle$ be an $N$-qubit state; then one requires that $E(|\psi\rangle|0\rangle)= E(|\psi\rangle)$. Here $|\psi\rangle|0\rangle$ is considered as an $(N+1)$-party state (and \emph{not} as an ancilla to one of the initial $N$ parties), where the $(N+1)$-th party is disentangled from the rest of the system.
\end{itemize}

Conditions P3 and P4 are often found in literature as necessary requirements for entanglement measures. Condition P5 has been introduced more recently \cite{universalityI}, in the context of the study of universality in MQC. Other different requirements have been  formulated, and for a more detailed analysis of them we refer to \cite{pleniovirmani}.

Depending on the set of conditions that are satisfied by the quantity $E$, we can define different types of measures. In particular, we can distinguish the following types, which we will use in the following sections.

\begin{defi} \
\begin{description}
\item[Weak entanglement monotone.] A real function $E$ is called a weak entanglement monotone if it satisfies conditions P1 to P3. \item[Strong entanglement monotone.] A real function $E$  is called a strong entanglement monotone if it satisfies conditions P1 to P4. \item[Extendable weak/strong monotone.] An extendable weak (strong) monotone is a weak (strong) entanglement monotone which additionally satisfies condition P5.
\end{description}
\end{defi}
Note that, in all these definitions, we are imposing the convexity of the function. This condition is not always deemed necessary, but the measures we consider in the following satisfy it. We also remark that every strong entanglement monotone is also a weak monotone. The notion of an extendable monotone was introduced in \cite{universalityI} under the name ``type II monotone''.

We now define another property, related to monotonicity under LOCC operations, that will be relevant in the analysis of resources for approximate measurement-based quantum computation.

\begin{itemize}
\item[P6.]{\it Weak non-increasing under LOCC}: a function $E$ is weakly non-increasing under LOCC if, for any state $\rho$ and for any LOCC protocol $\Lambda_\locc:\rho\to\{p_i,\rho_i\}$, we have $E(\rho)\geq\min_i E(\rho_i)$.
\end{itemize}
In other words, monotones satisfying P6 are such that at least one of the outputs of an LOCC protocol acting on an initial state $\rho$ has entanglement smaller than that of $\rho$. Such a condition is trivially satisfied by any strong entanglement monotone. We conjectured \cite{cat_thesis} that P6 is implied by weak monotonicity, but this still has not been proved.

To end this section, we will introduce two quantities associated to any entanglement measure $E$, which play a fundamental role both in \cite{universalityI} and in the results contained in Section \ref{sec:nogo}. The first notion is the \emph{asymptotic entanglement of a family of states}. Let $\Sigma=\{\sigma_i\}_i$ be an (infinitely large) family of many-qubit states, and $E$ be entanglement monotone defined on $N$-qubit states, for all $N$. We define the asymptotic entanglement $E(\Sigma)$ of the family as \beq E(\Sigma)=\sup_{\sigma\in\Sigma} E(\sigma). \eeq The case $E(\Sigma)=\infty$ is allowed.

Second, the \emph{asymptotic entanglement $E^*$ of $E$} is defined as
 \beq
 E^*=\sup_{\rho\in\mcs} E(\rho),
 \eeq
where the supremum is taken over all $N$-qubit states, for all $N\in\naturali$. The case $E^*=\infty$ is allowed. Note that, if $E$ is convex, one can restrict the set over which the supremum is taken to only the set of pure states (thus recovering the definition found in \cite{universalityI}).

\subsection{$\epsilon$-measures of entanglement}
\label{subsec:epsilonmeasures}

The $\epsilon$-monotones \cite{epsilon} are a class of entanglement monotones which can be associated to any existing monotone, and which depend on a precision parameter $\epsilon$. They have been introduced to address the issue of quantifying the entanglement contained in a state which is only partially known as in the case of, for example, a state prepared using an imperfect apparatus. Given any entanglement measure $E$, its $\epsilon$-version is defined as \beq \label{epsilon} E^{(D)}_\epsilon(\rho)=\min\{E(\sigma)~|~D(\sigma,\rho)\leq\epsilon\}, \eeq where $D$ is a distance on the set $\mcs$ of states which is convex and contractive under completely positive trace preserving maps
\cite{foot1}, and $\sigma,\rho\in\mcs$. To lighten notation we will often omit the superscript in ``$E^{(D)}_{\epsilon}$'' referring to the distance measure $D$ when writing down an $\epsilon$-measure, and we will simply write $E_{\epsilon}$.

The quantity $E_\epsilon$ quantifies the ``guaranteed'' entanglement contained in a state since, by definition, any state $\sigma$ within an $\epsilon$-distance of the desired state $\rho$ has entanglement $E(\sigma)\geq E_\epsilon(\rho)$. In the following we will see that the $\epsilon$-measure of a state is the crucial quantity to consider when studying approximate preparation of such a state. Indeed, if we aim at preparing a state which is $\epsilon$-close to $\rho$, then $E_\epsilon(\rho)$ is the minimum entanglement that we must be able to obtain from the initial resource state.

In the remainder of this section, we highlight some relevant properties of $\epsilon$-monotones.

First, it has been shown \cite{epsilon} that $E_\epsilon$ is always a weak entanglement monotone if $E$ is. Moreover, also property P5 illustrated above is inherited by the $\epsilon$-version of a monotone satisfying it. Therefore, the $\epsilon$-version of an extendable weak monotone is again an extendable weak monotone. On the other side, the $\epsilon$-version of an entanglement measure is never a strong monotone. We refer to \cite{epsilon} for details.

Computing the asymptotic entanglement $E_{\epsilon}^*$ for arbitrary $\epsilon$ may be a difficult task. Nevertheless, it is often tractable to compute the asymptotic entanglement $\Ee^*$ when we are interested in the limit $\epsilon\to0$. This is particularly true in the case of continuous measures, where the following observation holds true.

\begin{prop}
\label{prop:Estar} If $E$ is bounded (for any fixed dimension), convex, and continuous then $\lim_{\epsilon\to 0^+}\Ee^*=E^*$.
\end{prop}
\begin{proof}
Let $E^*\in(0,\infty]$. To prove the statement we have to show that, for any $\mu>0$, there exists $\bar{\epsilon}(\mu)>0$ such that $\epsilon\leq\bar{\epsilon}(\mu)\Rightarrow \Ee^*\geq E^*-\mu$.

Consider that, for any state $\rho$, we have that $\epsilon'\leq\epsilon\Rightarrow E_{\epsilon'}(\rho)\geq\Ee(\rho)$, which implies that $\epsilon'\leq\epsilon\Rightarrow E_{\epsilon'}^*\geq\Ee^*$. Moreover, from the definition of $\Ee^*$ it follows that, for any state $\rho$ and for any choice of $\epsilon$, $E^*_\epsilon\geq\Ee(\rho)$. This implies that, $\forall \epsilon \leq \bar{\epsilon}(\mu)$ and $\forall \rho$, we have $\Ee^*\geq E_{\bar{\epsilon}(\mu)}^*\geq E_{\bar{\epsilon}(\mu)}(\rho)$.

Therefore, it is sufficient to prove that \best \forall \mu>0~,~\exists \bar{\epsilon}(\mu),~\rho(\mu)\textrm{ such that }E_{\bar{\epsilon}(\mu)}(\rho(\mu))\geq E^*-\mu. \eest

In order to do so, we first recall that, since the family $\Psi_C=\{\ket{C_{N_i}}\}_i$ of two-dimensional cluster states (on $N_i=i\times i$ qubits) is exact and deterministic universal, we have that $E(\Psi_C)=E^*$, for any entanglement measure $E$ \cite{universalityI}. This implies that, for any $\mu>0$, there exists $N(\mu):=N_{i(\mu)}$ such that $E(\ket{C_{N(\mu)}})\geq E^*-\mu/2$.

In \cite{epsilon}, it has been shown that, if $E$ satisfies the hypotheses above, then $\Ee$ is continuous in $\epsilon$ and $\rho$. Hence, it is always possible to find an $\bar{\epsilon}(\mu,N(\mu))>0$ such that $E_{\bar{\epsilon}(\mu,N(\mu))}(\ket{C_{N(\mu)}})\geq E(\ket{C_{N(\mu)}})-\mu/2$.

We have thus that, for any $\mu>0$, there exists a state $\ket{C_{N(\mu)}}$ and an $\bar{\epsilon}(\mu,N(\mu))>0$ such that \best
\begin{split}
E_{\bar{\epsilon}(\mu,N(\mu))}^*&\geq E_{\bar{\epsilon}(\mu,N(\mu))}(\ket{C_{N(\mu)}})\\
&\geq E(\ket{C_{N(\mu)}})-\mu/2\geq  E^*-\mu.
\end{split}
\eest
\end{proof}

In the case of discontinuous measures, such as the $\chi$-width \cite{maarten1} or the Schmidt measure \cite{hans}, one has to compute $\Ee^*$ on a case by case basis. We will elaborate on the case of the $\chi$-width in section \ref{subsec:exmeasures}.

\subsection{Two entanglement measures}
\label{subsec:exmeasures}


In this Section we consider two explicit examples of entanglement measures that we use in Section \ref{sec:nogo} to construct criteria for approximate, non-deterministic universality. These are the geometric measure of entanglement and the Schmidt-rank width. We discuss in which sense these quantities are entanglement measures, what their asymptotic entanglement is, and how the $\epsilon$-versions of these measures behave.

\subsubsection{Geometric measure of entanglement}

The \emph{geometric measure} of entanglement was first introduced as a bipartite entanglement measure in \cite{shimony} and then generalized in \cite{barnumlinden, weigoldbart} to the multipartite setting. The intuition behind this measure is that the more entangled a state is, the more distinguishable it is from a separable state. The monotone can be defined as follows. Let $\spsi$ be a state of $N$ qubits, and let $\pi(\spsi)$ denote the maximum fidelity between $\spsi$ and a factorized state on $N$ qubits \beq \pi(\spsi)=\max_{\sphi=\ket{\phi_1}\otimes\cdots\otimes\ket{\phi_N}} |\langle\psi\sphi|^2. \eeq The geometric measure $E_G$ is defined by \beq E_G(\spsi) = 1- \pi(\spsi), \eeq This measure, defined for pure states, can be generalized to the case of mixed states by the convex roof construction, that is: \beq E_G(\rho)=\min_{\{p_i,\ket{\psi_i}\}_i}\sum_i p_i E_G(\ket{\psi_i}), \eeq where the minimum is taken over all $\{p_i,\ket{\psi_i}\}_i$ such that $\rho=\sum_i p_i\proj{\psi_i}$.

One can verify that such measure satisfies conditions P1 to P5 and is, thus, an extendable strong entanglement monotone (and therefore also an extendable weak monotone).

Next we consider the $\epsilon$-version of the geometric measure, and we focus on $\epsilon$-measures based on distances that are ``strictly related to the fidelity''.

\begin{defi}
\label{strictlyrelated} A distance $D$ on the set of states is said to be \emph{strictly related to the fidelity} if, for any two states $\rho$ and $\sigma$, $D(\rho,\sigma)\leq\epsilon \Rightarrow F(\rho,\sigma)\geq 1-\eta(\epsilon)$, with $0\leq\eta(\epsilon)\leq1$ a strictly monotonically increasing function of $\epsilon$ (for $\epsilon\geq0$ and $\epsilon$ less than the maximum value that $D$ can assume) such that $\eta(0)=0$.
\end{defi}
\noindent An example of such a measure is the trace distance.

The following is a technical result, which is a lower bound for $(E_G)_\epsilon(|\psi\rangle)$ in terms of $E_G(|\psi\rangle)$.

\begin{prop}
\label{cor:geom} Let $D$ be a distance measure that is strictly related to the fidelity. Further, let $(E_G)_\epsilon$ denote the corresponding $\epsilon$-geometric measure. Then, for any pure state $\spsi$ and for any choice of $\epsilon>0$ such that $\eta=\eta(\epsilon)\lesssim 0.44$, the quantity $(E_G)_\epsilon(\spsi)$ is not smaller than \beq
\begin{split}
\left[1-\left(\frac{3\sqrt{\eta}}{2 E_G(\spsi)}\right)^{2/3}\right]\left[E_G(\spsi)-(18E_G(\spsi)\eta)^{1/3}\right].
\end{split}
\eeq
\end{prop}
The proof of Proposition \ref{cor:geom} rather involved and will be given in Appendix \ref{sec:appendix_geo}.

The above result can be used to bound the asymptotic $\epsilon$-geometric entanglement $(E_G)_\epsilon^*$. We have:
\begin{prop}
\label{lemma:geomepsstar} Let $D$ be a distance measure that is strictly related to the fidelity, and let $\epsilon>0$ be such that $\eta(\epsilon)\leq 0.44$, where $\eta(\epsilon)$ is such that $D(\rho,\sigma)\leq\epsilon \Rightarrow F(\rho,\sigma)\geq1-\eta(\epsilon)$. If $(E_G)_\epsilon$ denotes the $\epsilon$-geometric measure with respect to distance $D$, then \beq (E_G)_\epsilon^*\geq 1- 4\eta^{1/3}+3.4\eta^{2/3} \eeq
\end{prop}
\begin{proof}
Since $(E_G)_\epsilon^*$ is defined as the supremum over all possible states, we have \best (E_G)_\epsilon^*\geq (E_G)_\epsilon(\Psi_C), \eest where $\Psi_C=\{\ket{C_{N_i}}\}_i$ is the family of two-dimensional cluster states on $N_i=i\times i$ qubits. The geometric measure for this class of states has been computed \cite{MMV07}, and we have $E_G(\ket{C_{N_i}})=1-2^{-N_i/2}$.

In order to prove the statement, we apply Proposition \ref{cor:geom} to obtain \beq
\begin{split}
(E_G)_\epsilon^* &\geq (E_G)_\epsilon(\Psi_C)\\
    &\geq \sup_N\left\{ \left[1-\left(\frac{3\sqrt{\eta}}{2 (1-2^{-N/2})}\right)^{2/3}\right]\right.\\
        &~~~~~~~~~~~~~\left.\left[1-2^{-N/2}-(18(1-2^{-N/2})\eta)^{1/3}\right]\right\}\\
     &= \left[1-\left(\frac{3\sqrt{\eta}}{2}\right)^{2/3}\right]\left[1-(18\eta)^{1/3}\right]\\
     &= 1-(\frac{9\eta}{4})^{1/3}-(18\eta)^{1/3}+(\frac{81}{2}\eta^2)^{1/3}\\
    &\geq 1 - 4\eta^{1/3}+3.4\eta^{2/3}.
\end{split}
\eeq
\end{proof}
Note that this result implies that \beq\lim_{\epsilon\to 0} (E_G)_\epsilon^* = 1.\eeq The latter also follows immediately from Proposition \ref{prop:Estar}.

\subsubsection{Schmidt-rank width}

The \emph{Schmidt-rank width} is an entanglement monotone which has been  introduced and investigated in \cite{maarten, maarten1, universalityI}.  It has been proved that this measure is an extendable strong entanglement monotone, and it can be used to assess whether resources for MQC admit an efficient classical simulation \cite{maarten1}.

The \emph{Schmidt-rank width} $\chi_\wid$ of a pure state $\spsi$ computes the minimum Schmidt rank $\chi$ of $\spsi$,  where the minimum is taken over a specific class of bipartitions of the system. More precisely, $\chi_\wid(\spsi)$ is defined as follows.
\begin{figure}
\begin{center}
\includegraphics[width=0.3\textwidth]{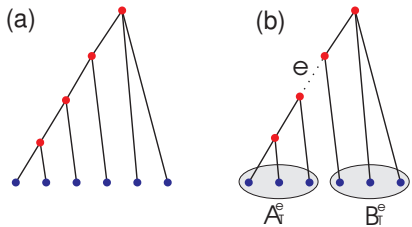}
\caption[Example of subcubic tree]{\footnotesize{(a) Example of a subcubic tree $T$ with six leaves. (b) Tree $T\backslash e$ obtained from $T$ by removing edge $e$, and induced bipartition.}}
\end{center}
\label{fig:subcubic}
\end{figure}
Let $\spsi$ be an $N$-partite state. We consider a subcubic tree $T$, i.e. a graph with no cycles, where each vertex has exactly 1 or 3 incident edges, with $N$ leaves ($N$ vertices with only 1 incident edge), which we identify with the $N$ parties of the system (see Figure \ref{fig:subcubic}). If $e=\{i,j\}$ is an arbitrary edge of $T$, we denote by $T\backslash e$ the graph obtained by deleting the edge $e$ from $T$. The graph then consists of two connected components, which naturally induce a bipartition $(A_T^e, B_T^e)$ of the system. If $\chi_{A_T^e,B_T^e}(\spsi)$  is the Schmidt rank of $\spsi$, with respect to the bipartition $(A_T^e, B_T^e)$, the \emph{Schmidt-rank width} of $\spsi$ is given by \beq \chi_\wid(\spsi)=\min_T\max_{e\in T} \chi_{A_T^e,B_T^e}(\spsi), \eeq where the minimum is taken over all subcubic trees $T$ with $N$ leaves (identified with the $N$ parties of the system), and $\chi_{A_T^e,B_T^e}(\spsi)$ is the Schmidt rank of $\spsi$ with respect to the bipartition $(A_T^e, B_T^e)$.

The Schmidt rank width may be generalized to mixed states by a convex roof construction.

Note that the Schmidt-rank width is not continuous, such that Proposition \ref{prop:Estar} cannot be used to compute the asymptotic behavior of its $\epsilon$-version in the limit $\epsilon\to0$. However, it is still relatively easy to gain insight in this matter, in the following way. First, note that \beq \label{E_wd} \chi_\wid(|\psi\rangle)\geq E_\wid(|\psi\rangle) \eeq for every state $|\psi\rangle$. Here $E_\wid(|\psi\rangle)$ denotes the entropic entanglement width, as defined in \cite{universalityI}. The entropic entanglement width is defined via the same optimization procedure as the Schmidt-rank width, now with the entanglement entropy as  the ``basic measure''. Note that (\ref{E_wd}) implies that \beq (\chi_\wid)_\epsilon(|\psi\rangle)\geq (E_\wid)_\epsilon(|\psi\rangle), \eeq and thus $(\chi_\wid)_\epsilon^*\geq (E_\wid)_\epsilon^*$.  Furthermore, as the entropic entanglement width is a weak monotone which is moreover continuous, and since $E_\wid^*=\infty$, one has  \beq (E_\wid)_\epsilon^* \xrightarrow{\epsilon\to0} E_\wid^*=\infty\eeq due to Proposition \ref{prop:Estar}. We can therefore conclude that also \beq \lim_{\epsilon\to0} (\chi_\wid)_\epsilon^*  =\infty. \eeq


\section{Universality in MQC}
\label{sec:universality}

In the one-way model of computation, information is processed by means of single qubit measurements on an initial highly entangled state. In the original proposal \cite{1-way1}, this state was chosen to be a cluster state, but there is no reason to assume that this is the only possible choice. Indeed, in recent works it has been shown that also other states can be used as a resource for measurement-based quantum computation \cite{universalityI,gross}.  Following \cite{universalityI}, in this work we consider the case in which any LOCC operation can be performed on the initial state. This corresponds to allowing two way classical communication, whereas the original scheme only requires one-way communication.

We report here the definition of universal CQ resources used in \cite{universalityI}, and on which the following discussion will be based.
\begin{defi}[Exact universal resources.]
A family $\Sigma=\{\sigma_i\}_i$ of states is called a universal resource for measurement-based quantum computation if, for every $N$ and for every $N$-qubit quantum state $\ket{\phi_\out}$, there exists an $M$-qubit resource state $\sigma\in\Sigma$ and an LOCC protocol $\Lambda_\locc$ that acts in the following way \beq \sigma\xrightarrow{\Lambda_\locc}P_\out \otimes P_0^{\otimes (M-N)}, \eeq where $P_\out=\proj{\phi_\out}$ and $P_0=\proj{0}$.
\end{defi}

\subsection{$\epsilon$-approximate $\delta$-stochastic universality }

While previous works have considered the characterization of exact universal resources for MQC, we are here more interested in considering weaker forms of universality, where the output state can be generated stochastically (with some finite success probability) or with some finite accuracy. Note that the nature of the resource might not be the only reason for which exact universality cannot be achieved. Indeed, as well as the circuit model with a finite gate  basis, one can consider the case of one-way quantum computation where there are, e.g., only finite possible measurement  directions \cite{foot2}. Also, one must consider the fact that any experimental implementation will introduce some source of error in the computation. In order to take these factors into account, in the following we define the concepts of $\delta$-stochastic and $\epsilon$-approximate universality. In a realistic scenario, one is expected to be interested mainly in approximate stochastic (or quasi-deterministic) universality.


\begin{defi}[$\epsilon$-approximate $\delta$-stochastic universal resources]
A family of states $\Sigma=\{\sigma_i\}_i$ is called $\epsilon$-approximate, relatively to a distance measure $D$, and $\delta$-stochastic universal if for every $N$ and for every $N$-qubit quantum state $|\phi_{\rm out}\rangle$, there exists an $M$ qubit state $\sigma \in \Sigma$ and an LOCC protocol with output branches $\{p_i,\rho_i\}$ such that the sum of the probabilities $p_i$ for the branches where $D_i=D(\rho_i,P_\out\otimes P_0^{\otimes M-N})\leq\epsilon$ (where $P_\out=\proj{\phi_\out}$ and $P_0=\proj{0}$) is larger than $1-\delta$.
\end{defi}

First, as regards the case of $\delta$-stochastic universality, we do not require that the output state is generated deterministically, but it is sufficient that this happens under stochastic LOCC (SLOCC) with sufficiently high probability, that is \best p_{\textrm{success}} = \sum_{i:\rho_i=P_\out\otimes P_0^{\otimes (M-N)}}p_i\geq 1- \delta . \eest In particular, when $\delta$ can be made arbitrary small, we may call this quasi-deterministic universality, which is stronger than $\delta$-stochastic universality for a fixed $\delta$ .



Second, as regards $\epsilon$-approximate universality, we require the output of the computation is generated approximately with accuracy $\epsilon$, as is the case for the quantum circuits built from a finite universal set of elementary gates. Precisely, $D$ can be any distance measure on the set of states, that is contractive under LOCC and convex. The choice of the appropriate measure might depend on the task for which the output state is required (see, for example, the related discussion in
\cite{nielsendistance}).



\subsection{Efficient universality}

We now  consider the issue of how to generalize the concept of efficient universality to the approximate and stochastic cases. In order to do so, let us first recall the definition of exact efficient universality \cite{universalityI}.
\begin{defi}[Exact efficient universal resources]
A family of states $\Sigma=\{\sigma_i\}_i$ is called an efficient exact universal resource for measurement-based quantum computation if, for every $N$ and for every $N$-qubit quantum state $\ket{\phi_\out}$ which can be obtained by a poly-sized quantum circuit, there exists an $M$-qubit state $\sigma\in\Sigma$, with $M\leq\mco(\poly(N))$, such that the transformation $\sigma\to\ket{\phi_\out}\ket{0}^{\otimes (M-N)}$ is possible by means of LOCC in time that is at most $\poly(N)$ and using classical processing that is polynomially bounded in space and time.
\end{defi}

This definition can be easily extended to the approximate and stochastic case, when the desired accuracy $\epsilon$ and success probability $\delta$ are fixed. In this case one has the following

\begin{defi}[Efficient  $\epsilon$-approximate $\delta$-stochastic universal resources]
Let $|\phi_\out\rangle$ be any $N$--qubit quantum state that can be generated efficiently, i.e. with a poly--sized quantum circuit, from a product state in the network model, and let $P_\out=\proj{\phi_\out}$. A family of states $\Sigma=\{\sigma_i\}_i$ is  efficient $\epsilon$-approximate (with respect to some distance $D$) $\delta$-stochastic universal if there exists an $M$-qubit state $\sigma\in\Sigma$, with $M\leq\mco(\poly(N))$, and an LOCC protocol with output branches $\{p_i,\rho_i\}_i$ such that \best \sum_{i:D(\rho_i,P_\out\otimes P_0^{\otimes(M-N)})\leq\epsilon}p_i\geq 1-\delta \eest (with $P_0=\proj{0}$) that can be implemented in $\mco(\poly(N))$ time, using classical side processing that is bounded in space and time by $\poly(N)$.
\end{defi}

For approximate and stochastic computation, in many cases it is meaningful and interesting to take into account also the scaling of the overhead with the desired accuracy $\epsilon$ and success probability $\delta$. In the circuit model, the scaling with the accuracy is determined by the Solovay-Kitaev theorem \cite{kitaev}. Similarly, in the one-way model we require that the scaling of the overhead in spatial, temporal and computational resources with $\epsilon$ is $O({\rm poly}(m,\log(1/\epsilon)))$ for states that can be produced with $m$ gates in the network model. Notice that we allow for a polynomial increase of resources with respect to the number of elementary gates $m$, as it is also done in the definition of {\em exact} efficient universality. It follows that any state that can be generated efficiently in the network model, i.e. with ${\rm poly}(m)$ elementary gates, should be approximated with accuracy $\epsilon$ with overhead that scales $O({\rm poly}(m,\log(1/\epsilon)))$ in the measurement-based model.

As regards the scaling with the probability parameter $\delta$, we claim that it should be treated in a way analogous to the accuracy, based on the following observation. Let us consider the following observation, in which we see that the two parameters $\delta$ and $\epsilon$ indeed play the same role when we try to determine the fidelity between the desired output of a computation on an $\epsilon$-approximate $\delta$-stochastic resource and the real output of the protocol.

\begin{obs}
Let us consider a universal $\epsilon$-approximate $\delta$-stochastic resource $\Sigma=\{\sigma_i\}_i$, and let $\Lambda_\locc$, such that $\sigma\to\{p_i,\rho_i\}_i$, be the LOCC protocol for some output $\ket{\phi_\out}$. We can, almost equivalently, consider $\Lambda_\locc$ to be performing the following transformation: $\sigma\to\rho=\sum_i p_i\rho_i$. Computing the fidelity between the desired output state $\ket{\phi_\out}$ and $\rho$, one finds that this leads to the bound $F(\rho,\ket{\phi_\out})\geq(1-\epsilon)(1-\delta)$.
\end{obs}

However, since the counterpart of the Solovay-Kitaev theorem for the success probability $\delta$ has not been found, it is not clear how we could attain an efficient scaling by ${\rm poly}(\log(1/\delta))$ in practice. Thus, we provide here a natural definition for efficient approximate and stochastic universal resources \cite{foot3}.

\begin{defi}[Efficient  approximate stochastic universal resources]
Let $|\phi_\out\rangle$ be any $N$--qubit quantum state that can be generated efficiently, i.e. with a poly--sized quantum circuit, from a product state in the network model, and let $P_\out=\proj{\phi_\out}$. A family of states $\Sigma=\{\sigma_i\}_i$ is  efficient approximate (with respect to some distance $D$) stochastic universal if, for all $\epsilon, \delta>0$, there exists an $M$-qubit state $\sigma\in\Sigma$, with $M\leq\mco(\poly(N,\frac{1}{\delta},\frac{1}{\epsilon}))$, and an LOCC protocol with output branches $\{p_i,\rho_i\}_i$ such that \best \sum_{i:D(\rho_i,P_\out\otimes P_0^{\otimes(M-N)})}p_i\geq 1-\delta \eest (with $P_0=\proj{0}$) that can be implemented in $\mco(\poly(N,\frac{1}{\delta},\frac{1}{\epsilon}))$ time, using classical side processing that is bounded in space and time by $\poly(N,\frac{1}{\delta},\frac{1}{\epsilon})$.
\end{defi}


\section{Criteria for universality and no-go results}
\label{sec:nogo}

In this Section we prove some necessary conditions for $\epsilon$-approximate $\delta$-stochastic universality, based on some entanglement properties of the resource. These results can be interpreted as a generalization of the ones
obtained in \cite{universalityI}, even though in some cases they require stronger assumptions on the entanglement monotone used to quantify the entanglement of the resource.

In \cite{universalityI} it was noticed that any deterministic exact universal resource $\Sigma$ must be such that, for any extendable entanglement measure $E$, $E(\Sigma)= E^*$. By evaluating $E^*$ in the case of different entanglement measures it was possible to show how some families of states (e.g. W states, 1-dimensional systems,...) could not be exact deterministic universal. In the case of $\epsilon$-approximate and $\delta$-stochastic universality, we show that a similar (but, naturally, weaker) result still holds true, where $E^*$ is substituted with $\Ee^*$. As we shall see in the following, though, in these more general cases it is necessary to consider entanglement measures $E$ satisfying some properties in addition to those required from the measures considered in  \cite{universalityI}.

While we are interested in the most general case of $\epsilon$-approximate and $\delta$-stochastic resources, we shall first treat the issue of $\epsilon$-approximate deterministic (i.e.  $\epsilon$-approximate and $\delta$-stochastic, with $\delta=0$) resources separately.

\subsection{$\epsilon$-approximate deterministic universality}

\begin{teo}[Criterion for $\epsilon$-approximate deterministic universality]
\label{teo:approxdet} Let $E$ be an extendable monotone that is weakly non-increasing under LOCC (as defined in Section \ref{subsec:axioms}), and let $\Sigma=\{\sigma_i\}_i$ be an $\epsilon$-approximate universal resource, with respect to some distance $D$. Then $E(\Sigma)\geq\Ee^{*}$. Furthermore, if $\Sigma$ is an approximate universal resource, then \beq E(\Sigma)\geq\lim_{\epsilon\to0^+}\Ee^{*}. \eeq
\end{teo}
\begin{proof}
Let us fix the distance measure $D$, let $\ket{\phi_\out}$ be any $N$-qubit state and $P_\out=\proj{\phi_\out}$. Since $\Sigma$ is $\epsilon$-approximate deterministic universal, there exist an $M$-qubit state $\sigma\in\Sigma$ and an LOCC protocol $\sigma\to\{p_i,\rho_i\}$ such that $D(\rho_i,P_\out) \leq\epsilon~\forall i$. Thus, for all $i$, \beq \label{eqapprdet1}
\begin{split}
E(\rho_i)&\geq \min_\rho\{E(\rho)| D(\rho,P_\out\otimes P_0^{\otimes(M-N)})\leq\epsilon\}\\
    &=\Ee(\ket{\phi_\out}\otimes\ket{0}^{\otimes{M-N}})\geq\Ee(\ket{\phi_\out}),
\end{split}
\eeq where $P_0=\proj{0}$ and in the last inequality we have used that, since $E$ is an extendable monotone, also $\Ee$ is \cite{epsilon}.

Since (\ref{eqapprdet1}) holds for all $\rho_i$, and we have assumed that $E$ is weakly non-increasing under LOCC, we have: \beq E(\sigma)\geq\min_i E(\rho_i)\geq\Ee(\ket{\phi_\out})~. \eeq The first part of the theorem is proved by considering the fact that $\ket{\phi_\out}$ is allowed to be any state. The second part of the theorem follows from the fact that $\Ee$, and thus $\Ee^*$, is monotonically non-increasing with $\epsilon$.

If $\Sigma$ is an approximate deterministic universal resource, then the previous result must hold true for any value of $\epsilon >0$.
\end{proof}

As we have mentioned above, computing $\Ee^*$ can in general be a hard task. Nevertheless, we have seen how this is possible at least in some particular cases. Whenever this happens, we can use Theorem \ref{teo:approxdet} to generalize the results obtained in the exact deterministic case also to the approximate (or even $\epsilon$-approximate) deterministic one, and show that some classes of states are not universal even in the approximate cases. These include, for examples, all graph states whose underlying graph has bounded  rank-width, such as, e.g.,  tree graphs or cycle graphs (which have bounded $\chi$-width)
\cite{maarten,maarten1}.

Moreover,  Proposition \ref{lemma:geomepsstar} also allows us to show that the family of $W$ states \cite{DurVC00-wstate} is not $\epsilon$-approximate universal for values of $\epsilon$ smaller than some finite $\bar\epsilon$.

\begin{es}
Let us consider the family $\Psi_W=\{\ket{W_N}\}_N$, where $\ket{W_N}$ is the $N$-qubit $W$ state \best \ket{W_N}=\frac{1}{\sqrt{N}}\sum_{i=1}^N\ket{e_{N,i}}, \eest and where $\ket{e_{N,i}}$ is defined to be the $N$-qubit computational basis state with a $\ket{1}$ in the $i$-th position, and $\ket{0}$ elsewhere. If $D$ is a distance measure that is strictly related to the fidelity (see Definition \ref{strictlyrelated}), then we have that $\Psi_W$ is not an $\epsilon$-approximate universal resource for any $\epsilon<\bar\epsilon$, where $\bar\epsilon$ depends on the choice of distance and is such that $\eta(\bar\epsilon)\simeq 0.1$\%, where $\eta$ is defined as in Definition \ref{strictlyrelated}.
\end{es}
\begin{proof}
If $\pi(W_N)$ is defined as in Section \ref{subsec:exmeasures}, we can consider $\pi(\Psi_W)=\sup_{W_N\in\Psi_W}\pi(W_N)$. Since it can be shown (c.f. Ref.~\cite{universalityI}) that $\pi(\Psi_W)=1/{\mathrm{e}}$, it follows that \best E_G(\Psi_W)=1-\frac{1}{\mathrm{e}}. \eest The statement follows immediately from  Proposition \ref{lemma:geomepsstar}, since we have that \beq E_G^{(\epsilon)^*}\geq 1- 4\eta^{1/3}+3.4\eta^{2/3}>1-1/{\mathrm{e}}=E_G(\Psi_W), \eeq for any choice of $\epsilon$ such that $\eta(\epsilon)\lesssim0.1\%$, and where $\eta=\eta(\epsilon)$ is such that $D(\rho,\sigma)\leq\epsilon \Rightarrow F(\rho,\sigma)\geq1-\eta(\epsilon)$.
\end{proof}

\subsection{$\epsilon$-approximate $\delta$-stochastic universality}

Let us consider, now, the case of $\epsilon$-approximate and $\delta$-stochastic universality. Also in this case we can formulate a criterion which generalizes the results obtained in \cite{universalityI} for exact deterministic universal resources, even though it is necessary to impose further requirements on the entanglement measure from which the criterion is derived.

\begin{teo}[Criterion for $\epsilon$-approximate $\delta$-stochastic universality]
\label{teo:apprstoc} Let $E$ be an extendable strong monotone, and let $\Sigma=\{\sigma_i\}_i$ be an $\epsilon$-approximate (with respect to a distance $D$) $\delta$-stochastic universal resource. Then \beq E(\Sigma)\geq(1-\delta)\Ee^*, \eeq where $\Ee$ is the $\epsilon$-generalization of $E$ with respect to $D$.
\end{teo}
\begin{proof}
Let us fix the distance measure $D$, let $\ket{\phi_\out}$ be an $N$-qubit quantum state and $P_\out=\proj{\phi_\out}$. Since $\Sigma$ is $\epsilon$-approximate deterministic universal, there exist an $M$-qubit state $\sigma\in\Sigma$ and an LOCC protocol $\sigma\to\{p_i,\rho_i\}$ such that \best \sum_{\epsilon-{\rm close}}p_i\geq(1-\delta), \eest where $P_0=\proj{0}$ and where the sum is taken over all indices $i$ such that $D(\rho_i,P_\out\otimes P_0^{\otimes(M-N)})\leq \epsilon$. We have then \best
\begin{split}
E(\sigma)&\geq\sum_i p_i E(\rho_i)\geq\sum_{\epsilon-{\rm close}}p_i E(\rho_i)\\
    &\geq \sum_{\epsilon-{\rm close}} p_i \min\{E(\rho)|D(P_\out\otimes P_0^{\otimes(M-N)},\rho)\leq\epsilon\}\\
    &= \sum_{\epsilon-{\rm close}} p_i \Ee(P_\out\otimes P_0^{\otimes(M-N)}) \\
    & \geq (1-\delta)\Ee(P_\out\otimes P_0^{\otimes(M-N)})\\
    &\geq(1-\delta)\Ee(\ket{\phi_\out}),
\end{split}
\eest where in the first inequality we have used the fact that $E$ is a strong monotone, and the last inequality follows from the fact that $E$ and, consequently, $\Ee$ are extendable monotones.
\end{proof}

An immediate consequence of this Theorem is the following
\begin{cor}
\label{cor:apprstoc} Let us consider an  strong monotone $E$, and let $\Ee$ be its $\epsilon$-generalization (with respect to some distance $D$) such that $\Ee^*=\infty$. Then any $\epsilon$-approximate $\delta$-stochastic universal family of resources $\Sigma$ is such that \best E(\Sigma)=\infty, \eest for all fixed values of $\delta<1$.
\end{cor}

Note that this implies that those families that were shown not to be approximate deterministic universal in the previous family are also not approximate $\delta$-stochastic universal, for all values of $\delta<1$.

\subsection{Efficiency in the approximate and stochastic case}

In the previous paragraphs, we have only considered criteria for universality, without taking efficiency into account. We will see now how these criteria can be strengthened to become necessary conditions for efficient $\epsilon$-approximate and $\delta$-stochastic universality. In order to do so, the strategy will be analogous to the one followed in \cite{universalityI} in the exact and deterministic case, and is based on the following observation:

{\bf Observation. }{\it A set of states $\Sigma=\{\sigma_i\}_i$ is an efficient $\epsilon$-approximate $\delta$-stochastic resource if and only if all 2-dimensional cluster states $\ket{C_{d\times d}}$ (for all $d$) can be prepared efficiently from the set $\Sigma$ by LOCC with success probability $p\geq(1-\delta)$ and with accuracy $\epsilon$. }
\begin{proof}
The necessity of the condition is immediate. The sufficiency follows from the fact that a family composed of states each of which is close to a cluster state is $\epsilon$-approximate $\delta$-stochastic universal for any choice of $\epsilon$ and $\delta$ (see Section \ref{subsec:approxresuniv}).
\end{proof}

As in the exact deterministic case, we see that the \emph{scaling} of entanglement plays a major role when one considers efficiency-related issues.

\begin{teo}[Criterion for efficient $\epsilon$-approximate $\delta$-stochastic universality]
\label{teo:efficientfam} Let $\Sigma=\{\sigma_i\}_i$ be an  $\epsilon$-approximate (with respect to some distance $D$) $\delta$-stochastic universal family, where $\sigma_i$ is a state on $N_i$ qubits. Let us consider an  extendable strong entanglement monotone $E$, and let $f_\epsilon$ be a function such that, for every 2-dimensional cluster state $\ket{C_{d\times d}}$ on $N=d^2$ qubits, one has \best \Ee(\ket{C_{d\times d}})\geq f_\epsilon(N), \eest
where $\Ee$ is the $\epsilon$-generalization of $E$ with respect to $D$.\\
If $E(\sigma_i)$ scales as $\log f_\epsilon(N_i)$, then $\Sigma$ cannot be an efficient $\epsilon$-approximate and $\delta$-stochastic universal resource.
\end{teo}
\begin{proof}
Since we have assumed that $\Sigma$ is an  $\epsilon$-approximate $\delta$-stochastic universal resource, for any $N=d^2$ there must exist a $g(N)$-qubit state $\sigma_{g(N)}\in\Sigma$ and an LOCC protocol $\sigma_{g(N)}\to\{p_i,\rho_i\}_i$ such that \best \sum_{\epsilon-{\rm close}}p_i\geq(1-\delta), \eest where the sum is taken over the indices $i$ such that $D(\sphii,\ket{C_{d\times d}}\ket{0}^{\otimes(g(N)-N)})\leq\epsilon$. From what we have already seen (see the proof of Theorem \ref{teo:apprstoc}), it follows that, necessarily \best E(\sigma_{g(N)})\geq(1-\delta)\Ee(\ket{C_{d\times d}})\geq(1-\delta)f_\epsilon(N). \eest In order for $\Sigma$ to be an efficient resource, though, it is necessary that $g(N)$ is at most polynomial in $N$, and thus, following an argument parallel to that in Theorem 9 of \cite{universalityI}, we can conclude that $E(\Sigma)$ cannot scale logarithmically with $f_\epsilon(N)$.
\end{proof}
We emphasize that the family of two-dimensional cluster states in principle does not play a distinguished role in Theorem  \ref{teo:efficientfam}, in the sense that it can be replaced -without weakening or strengthening the result- by any arbitrary efficient universal family or, in fact, any family of states which themselves also can efficiently be prepared.

\begin{es}
Based on the criterion by Theorem~\ref{teo:efficientfam}, the states whose Schmidt-rank width have a polylogarithmic scaling in $N$ are not efficient exact deterministic universal resources as shown in Ref.~\cite{universalityI}, nor efficient $\epsilon$-approximate and $\delta$-stochastic universal resources. These include the cluster state on the 2D stripe $ d \times \log d$, and the cluster state on the faulty 2D lattice with a site occupation probability $p \leq p_{c}$, as mentioned later in Sec.~\ref{sec:examples}.
\end{es}

\section{Examples of $\epsilon$-approximate and/or $\delta$-stochastic universal resources}
\label{sec:examples}

In this section we provide examples of families of states that are universal resource states when we relax our requirements for universal MQC to $\epsilon$-approximate and/or $\delta$-stochastic universality.

\subsection{2D cluster state with holes as an exact quasi-deterministic
resource}

Our model is a faulty 2D cluster state in which qubits get entangled after qubits are prepared with partial losses (called holes here) in the background 2D square lattice with total size $M = N^2$, where $N$ is the side length. The lattice-site occupation probability is denoted as $p_{\rm site}$, and thus the hole probability is given by $1-p_{\rm site}$. We assume here that every hole occurs independently according to the probability, and the locations of these holes are heralded. It is conceivable, for example in the implementations by optical lattice, that we may be able to check whether atoms are stored for each site before creating the 2D cluster state, and thus without destroying entanglement.

That is why, our faulty 2D cluster state with holes is considered to be {\it a pure graph state} corresponding to a specific configuration of holes, in contrast with the statistical ensemble (classical mixture of several configurations) characterized by $p_{\rm site}$. All statistical statements, such as the percolation phenomenon, are meant to hold true almost with certainty (more precisely, with probability approaching  unity in the thermodynamical limit), for all the possible realizations of the configuration of holes with a given $p_{\rm site}$.

\begin{es}[\cite{BEF+07}]
A family of 2D cluster states with holes (characterized by increasing total size $M$) is an efficient exact quasi-deterministic universal resource if and only if the site occupation probability $p_{\rm site}$ is greater than the percolation threshold $p_c = 0.5927 \ldots$ of the 2D square lattice.
\end{es}

\begin{proof}
The detailed proof is available in Ref.~\cite{BEF+07}, in which the phase transition of the computational power of the 2D cluster state with holes was proved at the above mentioned threshold $p_c$. See also a preceding work \cite{KRE07} for the use of percolation theory to prepare cluster states by non-deterministic gates.

In the supercritical phase ($p_{\rm site} > p_c$), it has been shown that if a preprocessing by polynomial-time classical computation is provided, we can construct an LOCC conversion which concentrates a perfect 2D cluster state from a faulty cluster state {\it with a constant overhead} (depending only on $p_{\rm site}$). Such an LOCC conversion works almost with certainty (namely, with success probability of LOCC conversion approaching  unity exponentially in $L$), and will produce the 2D cluster state with fidelity exactly one when it is available. That is why the resource is efficient exact quasi-deterministic universal.

\end{proof}

\subsection{Deformed 2D cluster state as an exact quasi-deterministic
resource} \label{subsec:deformed}

We now give an example of universal resources which is not a graph state. Let us consider a local deformation of the 2D $N\times N$ cluster state $|C_{N\times N}\rangle$,
\begin{equation}
|dC_{N\times N}\rangle = \left(\frac{2}{1+\lambda^2}\right)^{N^2/2} \Lambda^{\otimes N^2} |C_{N\times N}\rangle,
\end{equation}
where $\Lambda={\rm diag}(1,\lambda)$ is the local deformation parametrized by $\lambda$ such that, without loss of generality, $ 0 \leq \lambda \leq 1$. We call it a deformed 2D cluster state whereby the perfect 2D cluster state corresponds to $\lambda = 1$. The deformed 2D cluster state can be seen as a ``noisy'' 2D cluster state resulting probabilistically from the local filtering operation $\Lambda$. Note however that the fidelity with the perfect 2D cluster state is $\left(\frac{(1+\lambda)^2}{2(1+\lambda^2)}\right)^M$, i.e.,
 exponentially small in the number $M = N^2$ of the total qubits, so
that the inverse transformation to the perfect 2D cluster state (with the same size) will succeed only with an exponentially small probability. Nevertheless, we show that {\it one single copy} of such a system can be an efficient resource, regardless of its size $M$, when $\lambda$ lies above a certain threshold.

\begin{es}
A family of the 2D deformed cluster states (with the total size $M$ increasing) is an efficient exact quasi-deterministic universal resource if the deformation parameter $\lambda$ is larger than $0.6490 \ldots$ .
\end{es}

\begin{proof}
We show that one can convert the deformed cluster state $|dC_{N\times N}\rangle $ by means of LOCC {\it deterministically} into a graph state corresponding to a 2D $N\times N$ square lattice with holes. We apply local 2--outcome measurements described by POVM $\{\Lambda^{-1}={\rm diag}(\lambda,1),\ \overline{\Lambda^{-1}}={\rm diag}(\sqrt{1-\lambda^2},0)\}$ at each qubit. If the outcome $\Lambda^{-1}$ occurs, we successfully ``undo'' the effect of deformation, while when the outcome $\overline{\Lambda^{-1}}$ happens, the qubit is projected into $|0\rangle$ so that it corresponds to a deletion of the vertex with attached edges (i.e., a hole) in the 2D cluster state. The probability of these successful events, which is independent of the position of qubits, determines the site occupation probability,
\begin{equation}
p_{\rm site}=\frac{2\lambda^2}{1+\lambda^2}.
\end{equation}
 It should be noted that this expression is independent of the system size $M$.
According to the threshold $p_c$ of the 2D cluster state with holes, it is now clear that if $\lambda > \lambda_c \approx 0.6490 \ldots$ the resulting resource is efficient exact quasi-deterministic universal, so is true for the original deformed 2D cluster state. We remark that here $\lambda > 0.6490 \ldots$ is merely a sufficient condition for being efficiently universal.

\end{proof}


\subsection{A noisy cluster state as an $\epsilon$-approximate
deterministic resource} \label{subsec:approxresuniv}

\begin{es}
Let $\Sigma=\{\sigma_i\}_i$ be a family of mixed states such that, for all $i$, $\sigma_i=(1-p)\proj{C_{N_i}}+p\proj{\tilde C_{N_i}}$, where $\ket{C_{N_i}}$ is the 2-dimensional cluster state on $N_i=i\times i$ qubits, and $\ket{\tilde C_{N_i}}$ is obtained from $\ket{C_{N_i}}$ by applying a phase flip $\sigma_z$ on a single qubit, so that $\ket{\tilde C_{N_i}}$ has a $-1$ eigenvalue only at the corresponding stabilizer operator. Note that $p$ is independent of the total system size $N_i$,  because of the (unrealistic) assumption that only one phase flip can happen. Let $D$ be a convex distance measure on the set of states such that $D(\rho,\sigma)\leq 1$ for all $\rho$ and $\sigma$ \cite{foot4}. Then $\Sigma$ is an $\epsilon$-approximate deterministic universal resource, relatively to $D$, for $\epsilon\geq p$.
\end{es}
\begin{proof}
Let us consider any output state $\ket{\phi_\out}$ and let $P_\out$ be the projector onto such a state. Since the family of cluster states is exact and deterministic universal, then there exist a state $\ket{C_{N_i}}$ and an LOCC protocol that, acting on $\ket{C_{N_i}}$, generates the state $\ket{\phi_\out}$. This means that there exists an LOCC protocol $\Lambda_\locc$ such that $\Lambda_\locc[\proj{C_{N_i}}]=\sum_k p_k P_\out^{(A)} \otimes P_k^{(R)}$, where $P_k=\proj{k}$ are projectors onto orthogonal states of some register $R$. We have thus \best
\begin{split}
& \Lambda_\locc[\sigma_i] \\
&=(1-p)\Lambda_\locc[\proj{C_{N_i}}]+p\Lambda_\locc[\proj{\tilde C_{N_i}}]\\
&=(1-p)\sum_k p_k P_\out^{(A)}\otimes P_k^{(R)}+p\sum_k \tilde p_k \tau^{(A)}_k\otimes P_k^{(R)}
    \end{split}
\eest where $\Lambda_\locc[\proj{\tilde C_{N_i}}]=\sum_k \tilde p_k \tau^{(A)}_k\otimes P_k^{(R)}$. Since both $\ket{C_{N_i}}$ and $\ket{\tilde C_{N_i}}$ are 2-dimensional cluster states on $N_i$ qubits, we have that the probability of each output branch is the same and is
given by \cite{1way1long} \best p_k=\tilde p_k=\frac{1}{2^{N_i-m}}, \eest where $N_i-m$ is the number of qubits that are measured. We can thus write the final state of the system  plus the register as \best \Lambda_\locc[\sigma_i]=\sum_k \frac{1}{2^{N_i-m}} [(1-p)P_\out^{(A)}+ p \tau^{(A)}_k]\otimes P_k^{(A)}. \eest The $k$-th output branch, thus, yields a state $\rho_k=(1-p)P_\out+\tau^{(k)}$ such that \best
\begin{split}
D(\rho_k,P_\out)&=D((1-p)P_\out+p\tau^{(k)},P_\out)\\
    &\leq(1-p)D(P_\out,P_\out)+pD(\tau^{(k)},P_\out)\\
    &\leq p,
\end{split}
\eest where the first inequality derives from the convexity of the distance $D$, and the second one follows from the fact that $D(\rho,\sigma)\leq 1$. Since this holds for all the output branches, we obtain that the state $\ket{\phi_\out}$ has been produced $\epsilon$-approximately (for any $\epsilon\geq p$) and deterministically. The proof is completed by noticing that the argument holds for any desired output state $\ket{\phi_\out}$.
\end{proof}

We remark that a similar result holds not only for mixtures of two cluster states, but also for states of the form
\begin{equation}
\sigma_i = (1-p)\proj{C_{N_i}}+p \sum_{\bm k} \lambda_{\bm k} \proj{C_{N_i}^{\bm k}},
\end{equation}
where $\sum \lambda_{\bm k}=1$, ${\bm k}$ is a binary vector of length $N_i$ where $k_j \in \{0,1\}$ corresponds to qubit $j$, and $C_{N_i}^{\bm k}$ is a 2D cluster state which is obtained from  $\ket{C_{N_i}}$ by applying $(\sigma_z^{j})^{k_j}$ to qubit $j$, i.e. $|C_{N_i}^{\bm k}\rangle = \prod (\sigma_z^{j})^{k_j} \ket{C_{N_i}}$. Notice that the $\ket{C_{N_i}^{\bm k}}$ form a basis, and hence the noise term can also be the identity. Also the action of local Pauli noise channels acting on the individual qubits leads to states
of this form \cite{graphstatereview}. The key insight is again that the success probability for each branch is the same for all noise terms, leading to a distance $D(\rho_k,P) \leq p$ for the output states, independent of the measurement outcomes.

We also mention that a similar resource with the subsections~ \ref{subsec:deformed} and \ref{subsec:approxresuniv} has been considered recently in Ref.~\cite{BBD+08} through the analysis of the thermal state for the cluster-state Hamiltonian with a local $\sigma_{z}$ field.

\subsection{Stability of universal resources}

Let us consider a scenario in which one wants to experimentally implement some measurement-based computation. In this case, it is natural to assume that the initial resource cannot be prepared exactly. In the following Theorem~\ref{teo:resources}, we analyze this case, giving a proof of the stability of universal resources under initial perturbation, and determining an expression for the worsening of the probability and accuracy parameters as a function of the error in the initial preparation. Furthermore this also formally proves (taking into account the effect on both parameters $\epsilon$ and $\delta$) the intuitive idea that the computation on the approximate states can take place by means of the same LOCC protocol, thus the exact knowledge of the state is not necessary. This also implies that if computation on the original states was efficient, then it remains so also on the new states. Notice however that we do not consider here the case in which the LOCC protocol itself is faulty.

\begin{teo}
\label{teo:resources} Let $D$ be a convex, bounded  distance measure strictly related to the fidelity, such that the maximum distance between any two states be unity \cite{foot5}. Let us consider an (efficient) $\epsilon$-approximate (with respect to $D$) $\delta$-stochastic universal resource $\Gamma=\{\gamma_i\}_i$, with $\delta+\epsilon<1$. Moreover, let $\Sigma=\{\sigma_j\}_j$ be a family of states such that, for any $\gamma\in\Gamma$, there exists a state $\sigma\in\Sigma$ with $D(\sigma,\gamma)\leq\mu$ (for some $\mu\leq 1-\delta-\epsilon$). Then $\Sigma$ is an (efficient) $\epsilon'$-approximate $\delta'$-stochastic universal resource for any choice of $\epsilon'$ and $\delta'$ such that \best \delta' \eta(\epsilon')\geq\eta(\epsilon+\delta+\mu), \eest where $\eta(\epsilon)$ is such that $D(\rho,\sigma)\leq\epsilon\Rightarrow F(\rho,\sigma)\geq 1-\eta(\epsilon)$
\cite{foot6}.
\end{teo}

Its proof is given in Appendix \ref{appendixB}. Note that, in general, $\delta'$ and $\epsilon'$ will have to be (polynomially) larger than $\delta$ and $\epsilon$. If $D$ is the trace distance, then we have $\eta(\epsilon)\geq \epsilon$ thus obtaining that one can always find $\epsilon'$ and $\delta'$ satisfying the condition: \best \delta'\epsilon'\geq \epsilon+\delta+\mu. \eest Note that the condition we have found implies that $\delta'$ and $\epsilon'$ must be larger than, respectively, $\delta$ and $\epsilon$.

More importantly, though, Theorem \ref{teo:resources} implies that, whenever $\Gamma$ is an (efficient) deterministic exact universal resource, then one can choose any $\delta '$ and $\epsilon '$ such that $\delta '\epsilon '\geq\mu$. We have thus the following
\begin{cor}
Let $\Sigma=\{\sigma_i\}_i$ be an (efficient) exact deterministic universal resource and $D$ be any distance measure strictly related to fidelity. Then, for every $\delta, \epsilon>0$ there exists a $\mu>0$ such that any family $\tilde\Sigma=\{\tilde\sigma_i\}_i$ with $D(\sigma_i,\tilde\sigma_i)\leq \mu$, $\forall i$ is an (efficient) $\epsilon$-approximate (with respect to $D$) $\delta$-stochastic universal resource. Furthermore, if output $\ket{\phi_\out}$ is obtained by applying LOCC protocol $\Lambda_\locc$ on a state $\sigma_i\in\Sigma$, then the same protocol can be used on the corresponding state $\tilde\sigma_i\in\tilde\Sigma$ to produce an output that, with probability $p\geq(1-\delta)$, is within distance $\epsilon$ from $\ket{\phi_\out}$.
\end{cor}

This implies, in particular, that any family composed of states that are close enough to, e.g.,  a cluster state is $\epsilon$-approximate $\delta$-stochastic universal for some non-trivial choice of $\epsilon$ and $\delta$.


\section{Conclusions and outlook}
\label{sec:conclusions}

In this paper we have studied the issues of approximate and stochastic universality in measurement-based quantum computation. We have defined the concepts of approximate and stochastic universality, and shown how these concepts are not equivalent to each other by providing examples of resources that are approximate and deterministic universal, or exact stochastic universal. Generalizing the results obtained in  \cite{universalityI}, we have presented entanglement-based criteria that must be satisfied by any approximate (stochastic) universal resource. Moreover we have shown that such criteria are strong enough to allow us to discard some well-known families of states as non-universal, including e.g. GHZ states, W-states and 1D cluster states. The issue of efficiency has also been discussed, and we have shown how the previous results can be strengthened to include the request that a universal family of resources also allows for efficient computations. We found that entanglement needs to grow sufficiently fast for any approximate stochastic universal resource.

 On the other side, we have provided examples of resources that are approximate and/or stochastic universal. In particular, we have studied the case of a family of states that is only an approximation of some ($\epsilon$-approximate and/or $\delta$-stochastic) universal family. We have given a formal proof of the fact that such a family is always $\epsilon'$-approximate and $\delta'$-stochastic universal, and found an explicit bound for the scaling of the parameters $\epsilon'$ and $\delta'$ as functions of the original parameters $\epsilon$ and $\delta$, and of the degree of approximation of the family itself.  The proof also formalizes the intuitive idea that the computation on the approximate family can be performed by means of the same protocol that was devised for the exact family. In particular this means that if the initial resource was efficient universal, then also the approximate one is.

While we have found that basically any well behaved entanglement monotone can be used to obtain criteria for approximate and stochastic universality, one of the quantities  considered in \cite{universalityI}, the entropic entanglement width, does not fall under this category as it is not an entanglement monotone (in the terminology of \cite{universalityI}, more precisely, not a type-I monotone). For this measure it is not clear whether the results obtained for the exact, deterministic case can be lifted to the approximate, stochastic case. This affects in particular results about non-universality of states with a bounded or logarithmically growing block-wise entanglement, such as ground states of strongly correlated 1D quantum systems. We have also not touched the issue of encoded universality \cite{universalityI}, where the desired quantum states need only be generated in an encoded form. Also in this case it should be possible to obtain entanglement based criteria for approximate stochastic encoded universality, using the methods and techniques developed in this paper.

Finally, we would like to comment in relation to the results presented in \cite{GFE08,BMW08}, where it is shown  that a randomly chosen generic pure state (in other words the majority of all states) is no more useful as a resource for measurement-based quantum computation than a string of random {\it classical} bits, despite the fact that the former is colloquially often said to be almost maximally entangled.

Particularly related to the results presented in this paper, is the fact (proved in Ref.~\cite{GFE08}) that a family of states $|\psi_{M}\rangle$ on $M$ qubits, whose  geometric measure scales as $E_G(|\psi_{M}\rangle) \geq 1 - 2^{-M + {\mathcal O}(\log_2 M)}$ cannot provide a super-polynomial speed-up over classical computation with the aid of randomness and thus it is conceivably not a universal resource (unless the class of decision problems solvable by a probabilistic Turing machine in polynomial time with bounded error (BPP) coincides with the class of decision problems solvable by a quantum computer in polynomial time with bounded error (BQP)).

Note that it is required that the scaling of the geometric measure is even faster (by a constant factor in the front of $M$ in the exponent) than that of the cluster state for any spatial dimension, $E_G(|C_M\rangle) = 1 - 2^{-\lfloor M/2 \rfloor}$ \cite{MMV07}, and thus these states $|\psi_{M}\rangle$ can be considered highly entangled with regards to this measure (in the sense that such a family would not fail the criterion for universality based on the geometric measure).

There are two kinds of examples in Ref.~\cite{GFE08} which are shown to have such a scaling of the geometric measure. The first example is given by generic Haar-random pure states. It is not clear for us whether they also pass the necessary conditions illustrated in the previous sections if one considers other entanglement measures, although it is possible. However, it should be noted that these states already inherit ``unphysical'' complexity as resource states since it might not be possible to prepare them in a time polynomial  in $M$.

The second, efficiently preparable, example is given by a tree tensor network state. While in \cite{GFE08} it is shown that these states have indeed high geometric measure, it should be noted that its Schmidt-rank width is bounded without reaching the maximum (because of the constant tree width \cite{foot7}). We could therefore interpret that its uselessness (as a universal resource for MQC) originates from being {\it too little} entangled in terms of the Schmidt-rank width: the family would in fact fail the criteria illustrated in the previous sections when one bases them on this entanglement measure.

It would be interesting to see whether it is possible to find necessary criteria such as the ones shown in this work that allows us to discard random pure states  (and some pseudo random pure states which are efficiently preparable in case they are not universal either (cf.~\cite{Low09})) as non-universal. It is possible that such states already fail the criteria for some existing entanglement measure (other than the geometric measure), but it might prove necessary to identify a new one in order to obtain this result. We note that randomness in the description of the resource does not necessarily taint its usefulness immediately, as can be seen for instance by our Example~3.

\section*{Acknowledgements}

C.M. and M.P. thank M. Bremner and B. Kraus for discussions. A.M. acknowledges helpful discussions about Refs.~\cite{GFE08,BMW08} with D. Gross, J. Eisert, S. Flammia, and Z. Ji. We acknowledge support by the Austrian Science Fund (FWF), in particular through the Lise Meitner Program (M.P.), and the EU (OLAQUI,SCALA,QICS). The research at the Perimeter Institute is supported by the Government of Canada through Industry Canada and by Ontario-MRI.

\appendix

\section{Proof of Proposition \ref{cor:geom}}
\label{sec:appendix_geo}

In order to prove Proposition \ref{cor:geom}, we will first prove the following result.
\begin{prop}
\label{teo:geom} Let $D$ be a distance measure on the set of states that is strictly related to the fidelity, and let $E_G^{(\epsilon)}$ denote the corresponding $\epsilon$-geometric measure. Then, for any pure state $\spsi$ and for any $\epsilon>0$, \beq \label{eq:Teo1}
\begin{split}
E_G(\spsi) &\geq (E_G)_\epsilon(\spsi) \\
    & \geq \max_{\Delta>0} \left(1-\frac{\eta(\epsilon)}{\Delta}\right)\left(E_G(\spsi)-3\sqrt{\Delta}\right)
\end{split}
\eeq where $\eta(\epsilon)$ is such that $D(\rho,\sigma)\leq\epsilon \Rightarrow F(\rho,\sigma)\geq1-\eta(\epsilon)$.
\end{prop}

This result will yield the proof of Proposition \ref{cor:geom}. Indeed, let us consider the right-hand-side of the inequality \eqref{eq:Teo1}. We have that \best
\begin{split}
(E_G)_\epsilon(\spsi) &\geq \max_{\Delta>0} \left(1-\frac{\eta}{\Delta}\right)\left(E_G(\spsi)-3\sqrt{\Delta}\right)\\
    &= E_G(\spsi) +\max_{\Delta>0}(3\eta/\sqrt{\Delta}-3\sqrt{\Delta}-E_G(\spsi)\eta/\Delta)\\
    &\geq E_G(\spsi) -\min_{\Delta>0}(3\sqrt{\Delta}+E_G(\spsi)\eta/\Delta).
\end{split}
\eest Proposition \ref{cor:geom} is then proved straightforwardly by showing that, for $\eta\leq0.44$ the above minimum is obtained by $\Delta=\left(\frac{2}{3}E_G(\spsi)\eta\right)^{2/3}$.

We now prove Proposition \ref{teo:geom}. In order to do this we will need the following two lemmas.
\begin{lemma}
\label{lemmamixed} Let $\rho$ be a mixed state. Then, for any decomposition $\rho=\sum_i p_i \proj{\psi_i}$ of $\rho$ into pure states, one has that \beq \langle\psi|\rho|\psi\rangle\geq1-\eta \Rightarrow\sum_{i:|\langle\psi_i|\psi\rangle|^2\geq 1-\Delta}p_i\geq 1-\frac{\eta}{\Delta}, \eeq for any choice of $\eta\in[0,1]$ and $\Delta> 0$.
\end{lemma}
\begin{proof}
The statement is proved as follows. \best
\begin{split}
1-\eta\leq\bra{\psi}\rho\ket{\psi}&=\sum_i p_i |\langle\psi_i\ket{\psi}|^2\\
    &=\sum_{i:\Delta-{\textrm{close}}} p_i |\langle\psi_i\ket{\psi}|^2 + \sum_{i:\Delta-{\textrm{far}}} p_i |\langle\psi_i\ket{\psi}|^2\\
&\leq \sum_{i:\Delta-{\textrm{close}}} p_i + (1-\Delta)\sum_{i:\Delta-{\textrm{far}}} p_i \\
    &= \sum_{i:\Delta-{\textrm{close}}} p_i+(1-\Delta)\left(1-\sum_{i:\Delta-{\textrm{close}}} p_i\right)\\
&=1-\Delta+\Delta\sum_{i:\Delta-{\textrm{close}}} p_i
\end{split}
\eest where by ``close'' and ``far'' we refer respectively to those values of the indices $i$ such that $|\langle\psi_i|\psi\rangle|^2\geq1-\Delta$ or $ |\langle\psi_i|\psi\rangle|^2< 1-\Delta$  respectively.
\end{proof}

\begin{lemma}
\label{lemmaEGpure} For any two pure states $\ket{\psi}$ and $\ket{\tilde\psi}$ such that $|\langle\psi|\tilde\psi\rangle|^2\geq 1-\eta$ the following holds: \best \left\vert E_G(\psi)-E_G(\tilde\psi)\right\vert\leq 3\sqrt{\eta} \eest
\end{lemma}
\begin{proof}
Let $\ket{\psi}$ and $\ket{\tilde\psi}$ be two pure states such that $|\langle\psi|\tilde\psi\rangle|^2\geq 1-\eta$, and let $\ket{\Phi}$ and $\ket{\tilde\Phi}$ be completely factorized pure states such that $|\langle\psi|\Phi\rangle|^2=\pi(\psi)$ and $|\langle\tilde\psi|\tilde\Phi\rangle|^2=\pi(\tilde\psi)$. Then, \best
\begin{split}
\sqrt{1-\pi(\tilde\psi)}& =D_\tr(\ket{\tilde\psi},\ket{\tilde\phi}) \leq D_\tr(\ket{\tilde\psi},\ket{\phi}) \\
    &\leq D_\tr(\ket{\tilde\psi,\ket{\psi}})+D_\tr(\ket{\psi},\ket{\phi})\\
    & \leq\sqrt{\eta}+\sqrt{1-\pi(\psi)},
\end{split}
\eest where the equality follow from the properties of the trace distance for pure states, and the inequalities follow respectively from the fact that $\ket{\tilde\phi}$ is the pure state with minimum distance from $\ket{\psi}$, the triangle inequality for the trace distance, and the hypotheses. By squaring both sides of the inequality, we obtain \best 1-\pi(\tilde\psi)\leq\eta+1-\pi(\psi)+2\sqrt{\eta}\sqrt{1-\pi(\psi)}\leq 1-\pi(\psi) +3\sqrt{\eta}. \eest The same argument can be repeated inverting the roles of $\tilde\psi$ and $\psi$, thus obtaining the relation $|\pi(\psi)-\pi(\tilde\psi)|\leq3\sqrt{\eta}$, which implies \best
\begin{split}
\left\vert E_G(\psi)-E_G(\tilde\psi)\right\vert&=\left\vert 1-\pi(\psi)-1+\pi(\tilde\psi)\right\vert\\
    &=|\pi(\psi)-\pi(\tilde\psi)|\leq3\sqrt{\eta}.
\end{split}
\eest
\end{proof}

We can now proceed to proving Proposition \ref{teo:geom}.

\begin{proof}
Let $\spsi$ be a pure state, and let $\rho$ be such that $E_G(\rho)=E_G^{(\epsilon)}(\spsi)$. Furthermore, let us assume to have taken an optimal decomposition $\rho=\sum_i p_i \proj{\psi_i}$ into mixed states such that $E_G(\rho)=\sum_i p_i E_G(\spsii)$. Then, for any $\Delta>0$, we have \beq
\begin{split}
E_G(\rho)&=\sum_i p_i E_G(\spsii)\geq\sum_{i:\Delta-{\textrm{close}}}p_i E_G(\spsii) \\
&\geq \sum_{i:\Delta-{\textrm{close}}}p_i [E_G(\spsi) -3\sqrt{\Delta}]\\
&\geq \left(1-\frac{\eta(\epsilon)}{\Delta}\right) [E_G(\spsi)-3\sqrt{\Delta}]
\end{split}
\eeq where the second inequality derives from Lemma \ref{lemmaEGpure}, the last one from Lemma \ref{lemmamixed}, and $\eta(\epsilon)$ is such that $D(\rho,\sigma)\leq\epsilon \Rightarrow F(\rho,\sigma)\geq1-\eta(\epsilon)$.
\end{proof}

%





\section{Proof of Theorem~\ref{teo:resources}}
\label{appendixB}

\begin{proof}
Let $\ket{\phi_\out}$ be any desired $N$-qubit output state, and let $P_\out=\proj{\phi_\out}$. Since $\Gamma$ is an $\epsilon$-approximate $\delta$-stochastic universal resource, there exist an $M$-qubit state $\gamma\in\Gamma$ and an LOCC protocol acting on $\gamma$ with output $\{p_i,\rho_i\}_i$ such that \beq \label{eq:apprresproof} \sum_{\epsilon-{\rm close}}p_i\geq 1-\delta. \eeq where the sum is taken over all indices $i$ such that $D(\rho_i,P_\out\otimes P_0^{\otimes (M-N)})\leq\epsilon$, and $P_0=\proj{0}$.
This means that there exists an LOCC protocol $\Lambda_\locc$ such that $\rho_{AR}=\Lambda_\locc[\gamma]=\sum_i p_i \rho_i^{(A)}\otimes P_i^{(R)}$, where $P_i$ are projectors onto orthogonal states of some register $R$. We define $\rho_A=\tr_R(\rho_{AR})=\sum_i p_i \rho_i^{(A)}$.\\
Let $\sigma\in\Sigma$ be such that $D(\sigma,\gamma)\leq\mu$ and let us define $\tilde\rho_{AR}=\Lambda_\locc[\sigma]$ and $\tilde\rho_A=\tr_R(\tilde\rho_{AR})$.\\
Since $D$ is contractive under CPT maps we have \best D(\tilde\rho_A,\rho_A)\leq D(\tilde\rho_{AR},\rho_{AR})\leq D(\sigma,\gamma)\leq\mu . \eest This implies that \best
\begin{split}
D(\tilde\rho_A,&P_\out\otimes P_0^{(M-N)})\\
    &\leq D(\tilde\rho_A,\rho_A)+D(\rho_A,P_\out\otimes P_0^{(M-N)})\\
    &\leq\mu+D(\rho_A,P_\out\otimes P_0^{(M-N)}),
\end{split}
\eest
where we have used the triangle inequality in the second step.\\
Since $D$ is convex, $D(\rho_A,P_\out\otimes P_0^{(M-N)})\leq\sum_i p_i D(\rho_i,P_\out\otimes P_0^{(M-N)})$, thus \best
\begin{split}
D(\rho_A,&P_\out\otimes P_0^{(M-N)})\leq \sum_i p_i D(\rho_i,P_\out\otimes P_0^{(M-N)}) \\
    &= \sum_{\epsilon-{\rm close}} p_i D(\rho_i,P_\out\otimes P_0^{(M-N)} )\\
    &+ \sum_{\epsilon-{\rm far}}p_i D(\rho_i,P_\out\otimes P_0^{(M-N)} )\\
    &\leq \epsilon \sum_{\epsilon-{\rm close}} p_i +\sum_{\epsilon-{\rm far}}p_i\\
        &\leq \epsilon+\delta,
\end{split}
\eest where the sum over $\epsilon$-close and $\epsilon$-far are taken over the indices $i$ such that $D(\rho_i,P_\out\otimes P_0^{(M-N)})\leq\epsilon$ and $D(\rho_i,P_\out\otimes P_0^{(M-N)})>\epsilon$ respectively. In the last inequality we have used the fact that $\sum_{\epsilon-\textrm{close}} p_i\leq 1$, and that Eq. (24) holds true.

We have thus \beq D(\tilde\rho_A,P_\out\otimes P_0^{\otimes (M-N)})\leq\mu+\epsilon+\delta. \eeq Since $D$ is strictly related to fidelity, this implies that $F(\rho_A,P_\out\otimes P_0^{\otimes (M-N)})\geq 1-\eta(\mu+\epsilon+\delta)$. Defining two parameters $\delta', \epsilon'$ such that $1-\eta(\mu+\epsilon+\delta) :=1-\delta'\eta(\epsilon')$, and applying Lemma \ref{lemmamixed}, we obtain \best
\begin{split}
1-\delta'&\leq \sum_{i:|\langle\psi_i\ket{\phi_\out}\ket{0}^{\otimes (M-N)}|^2\geq 1- \eta(\epsilon')}q_i \\
    &=\sum_{i:D(\sphii,\ket{\phi_\out}\ket{0}^{\otimes (M-N)})\leq\epsilon'} q_i,
\end{split}
\eest
for any decomposition of $\tilde\rho_A$ into pure states $\tilde\rho_A=\sum_i q_i \proj{\psi_i}$. \\
Noticing that the argument holds for any output state $\ket{\phi_\out}$ completes the proof.
\end{proof}

\bibliographystyle{alpha}

\begin{thebibliography}{VMDB06}
\bibitem{foot1}
A distance $D$ on the set of states is said to be contractive under completely positive and trace preserving (CPTP) maps if, given any CPTP map $\Lambda$, then $D(\rho,\sigma)\geq D(\Lambda[\rho],\Lambda[\sigma])$, for any two states $\rho$ and $\sigma$.
\bibitem{foot2}
One can show that the model is $\epsilon$-approximate deterministic universal.
\bibitem{foot3}
One could demand the polylogarithmic scaling for $1/\delta$ and $1/\epsilon$ in the definition.
\bibitem{foot4}
Note that any bounded distance measure $D$ can be re-scaled so that it satisfies the condition $D(\rho,\sigma)\leq 1$.
\bibitem{foot5}
Note that, since the distance is chosen to be bounded, it is always possible to re-scale it so that the maximum distance between two states is 1.
\bibitem{foot6}
The statement can be easily generalized to the case in which the distance is bounded by some constant $\bar{D}\neq 1$. In this case, one requires $\epsilon+\delta+\mu<\bar{D}$, and the bound becomes $\delta' \eta(\epsilon')\geq\eta(\epsilon+\bar{D}\delta+\mu)$.
\bibitem{foot7}
The resource is a tree and thus its tree width is constant if we consider the local dimension is $d$, which actually scales as $\sqrt{\log M}$ in the construction of Ref.~\cite{GFE08}. Accordingly, the entanglement-based criterion for {\it encoded} universality should be applied, strictly speaking.
\end{thebibliography}

\newcommand{\etalchar}[1]{$^{#1}$}

\end{document}